\documentclass[iop, revtex4]{emulateapj}

\usepackage[pdftex]{lscape}

\slugcomment{}

\shorttitle{Spectroscopic observations of Ly$\alpha$ emitters at $z\sim7.7$}
\shortauthors{A. L. Faisst et al.}

\begin{document}


\title{Spectroscopic observation of Ly$\alpha$ emitters at $z\sim7.7$ and implications on re-ionization}


\author{A. L. Faisst\altaffilmark{1}, P. Capak\altaffilmark{2}, C. M. Carollo\altaffilmark{1}, C. Scarlata\altaffilmark{3}, N. Scoville\altaffilmark{4}}

\affil{\textit{Accepted April 25, 2014}}



\altaffiltext{1}{Institute for Astronomy, Swiss Federal Institute of Technology (ETH Zurich), CH-8093 Zurich Switzerland}
\altaffiltext{2}{Spitzer Science Center, 314-6 Caltech, Pasadena, CA, 91125}
\altaffiltext{3}{Minnesota Institute for Astrophysics, University of Minnesota, Minneapolis, MN 55455, USA}
\altaffiltext{4}{California Institute of Technology, 314-6 Caltech, Pasadena, CA, 91125}



\begin{abstract}
We present spectroscopic follow-up observations on two bright Ly$\alpha$ emitter (LAE) candidates originally found by \citet{KRUG12} at a redshift of $z \sim 7.7$ using the Multi-Object Spectrometer for Infra-Red Exploration (MOSFIRE) at Keck. We rule out any line emission at the $>5\sigma$ level for both objects, putting on solid ground a previous null result for one of the objects. The limits inferred from the non-detections rule out the previous claim of no or even reversed evolution between $5.7 < z < 7.7$ in the Ly$\alpha$ luminosity function (LF) and suggest a drop in the Ly$\alpha$ luminosity function consistent with that seen in Lyman Break galaxy (LBG) samples.  We model the redshift evolution of the LAE LF using the LBG UV continuum LF and the observed rest-frame equivalent width distribution.
From the comparison of our empirical model with the observed LAE distribution, we estimate lower limits of the neutral hydrogen fraction to be 50~-~70\% at $z~\sim~7.7$.
Together with this, we find a strong evolution in the Ly$\alpha$ optical depth characterized by $(1+z)^{2.2\pm0.5}$ beyond $z=6$ indicative of a strong evolution of the IGM.
Finally, we extrapolate the LAE LF to $z\sim9$ using our model and show that it is unlikely that large area surveys like UltraVISTA or Euclid pick up LAEs at this redshift assuming the current depths and area.
\end{abstract}





\section{Introduction}

Understanding when and how the universe re-ionized is fundamental to our understanding of how galaxies and large scale structure form and evolve and is sensitive to global cosmological parameters. In particular, the \textit{fraction of neutral hydrogen}, $x_{HI}$, in the intergalactic medium (IGM) is closely tied to early galaxy formation because it is related to the gas accretion rate onto galaxies. From current measurements it is still unclear when re-ionization occurred and what the sources of re-ionizing radiation are.

The best of such current measurements come from cosmic microwave background (CMB) experiments and high-redshift quasar studies, with additional constraints from Lyman Break (LBG) and Ly$\alpha$ emitting (LAE) galaxy studies.
WMAP \citep{LARSON11} and Planck \citep{TAUBER10} place a $\sim2-3\sigma$ constraint on when re-ionization occurred, based on the optical depth to the CMB due to Thompson scattering of electrons. These data are usually fit by a quick re-ionization at $z\sim10.5$, but are also fully consistent with a more gradual re-ionization with a tail ending at $z\sim6-7$ \citep{KOMATSU11,PLANCK13}.
Direct measurements of the optical depth from quasars indicate that the universe is neutral up to $z\sim7.1$, based on the highest redshift quasars known today \citep{FAN06,MCGREER11,MORTLOCK11}.
Furthermore, ultraviolet (UV) continuum measurements of LBGs between $z\sim7-10$ \citep{BOUWENS11,BRADLEY12,SCHENKER12,MCLURE13} suggest that galaxies have a difficult time re-ionizing the universe until later times unless the luminosity function is unusually steep at the faint end, or the continuum escape fraction is high \citep{ROBERTSON13}.

The fraction of strong Ly$\alpha$ emitters within LBG samples should give us a more direct, complementary, and unique measurement of $x_{HI}$ and therefore how quickly and when the universe is re-ionizing.

Fundamentally, Ly$\alpha$ photons are scattered in areas where the IGM contains more neutral hydrogen, so the escape fraction of Ly$\alpha$ photons is proportional to the volume of re-ionized hydrogen around the young galaxies. Hence the fraction of galaxies with strong  Ly$\alpha$ emission is related to the neutral fraction of the IGM \citep{HAIMAN99,MALHOTRA04,DIJKSTRA07,MALHOTRA06, DIJKSTRA10}.  However, it is important to note that this probe is also sensitive to the evolution of the interstellar medium (ISM) inside galaxies (like dust, see \citet{BOUWENS12b,FINKELSTEIN12,MALLERY12}), so one must understand the effects of galaxy evolution to probe the IGM.

The Ly$\alpha$ emission of LBG galaxies (selected using broad bands) is indicative of re-ionization ending at $z\sim6-7$ and a neutral hydrogen fraction of $\sim50$\% at $z\sim7$ \citep{FONTANA10, STARK10,PENTERICCI11,ONO11,SCHENKER11,CARUANA13}. In particular the fraction of strong Ly$\alpha$ emitters in LBG samples is found to rapidly drop beyond $z > 6.5$ over a range of $\Delta z \gtrsim 1$, a timescale of only $\sim 200$ Myrs \citep{STARK11,CURTISLAKE12,SCHENKER11}.

An alternative to LBG selection is the use of narrow-band (NB) filters to directly detect LAEs at specific redshifts (e.g., \citet{MALHOTRA01,HU04} and references therein). This method allows one to directly map the Ly$\alpha$ LFs as a function of redshift, which can then be compared to the LBG UV continuum LFs to estimate the neutral IGM fraction.

An overall change in the Ly$\alpha$ LFs between $5.7 < z < 6.6$ has been firmly established by large samples of spectroscopically confirmed LAEs \citep{OUCHI08,OUCHI10, HU10, KASHIKAWA11,MALHOTRA04}.
But the source of this change could be either an evolution in the IGM or a change in the internal ISM of the galaxies.

The evolution of the Ly$\alpha$ LF based on LAEs beyond $z>7$ is far less clear.
Apart from a few spectroscopically confirmed LAEs at $z\sim7$ (one spectroscopically confirmed out of two at $z=6.96$ \citep{OTA08} and one spectroscopically confirmed out of three at $z=7.22$ \citep{SHIBUYA12}), there are no confirmed LAEs at higher redshifts.
A total sample of $\sim15$ candidate LAEs at $z=7.7$ is known \citep{HIBON10,TILVI10,KRUG12}.
\citet{TILVI10} and \citet{KRUG12} favor a non-evolution of the Ly$\alpha$ LF between $5.7 < z < 7.7$ (see also \citet{HIBON11}), which is in tension with other narrow band searches for LAEs at $z > 7$ that only place limits on the number counts of LAEs \citep{SOBRAL09,CLEMENT12,OTA12,MATTHEE14}.
The reason for this tensions may be low-redshift interlopers and false detections in the LAE samples. At $z<7$, both of these are estimated to contribute less than 10--20\% (see e.g., \citet{OUCHI10}), at higher redshifts, these contribution are not known, yet, but are probably much higher (see \citet{MATTHEE14} and this work).
Spectroscopic follow-up observations of high redshift candidate LAEs are therefore necessary to resolve the tensions between the LAE and LBG results at $z>7$ and to constrain the process of re-ionization at higher redshifts.

In this work\footnote{Magnitudes are given in the AB system and we assume a flat universe with $\Omega_{m}=0.25$, $\Omega_{\Lambda}=0.75$, and H$_{0}=70$ km s$^{-1}$ Mpc$^{-1}$.}, we present Keck-I MOSFIRE spectroscopic follow-up of two $z\sim7.7$ LAE candidates ($\S2$) originally found by \citet{KRUG12}. 
We then go on to compare these results to existing data at lower redshift ($\S3$) and to an empirical model derived from the LBG UV continuum LF and observed equivalent width distribution ($\S\S4.1,4.2$).
This allows us to place limits on the neutral fraction of the IGM at $z\sim8$ ($\S4.3$) and enables us to predict the LAE LF at $z\sim8-9$ ($\S5$).


\begin{figure*}
\includegraphics[scale=.25,angle=0]{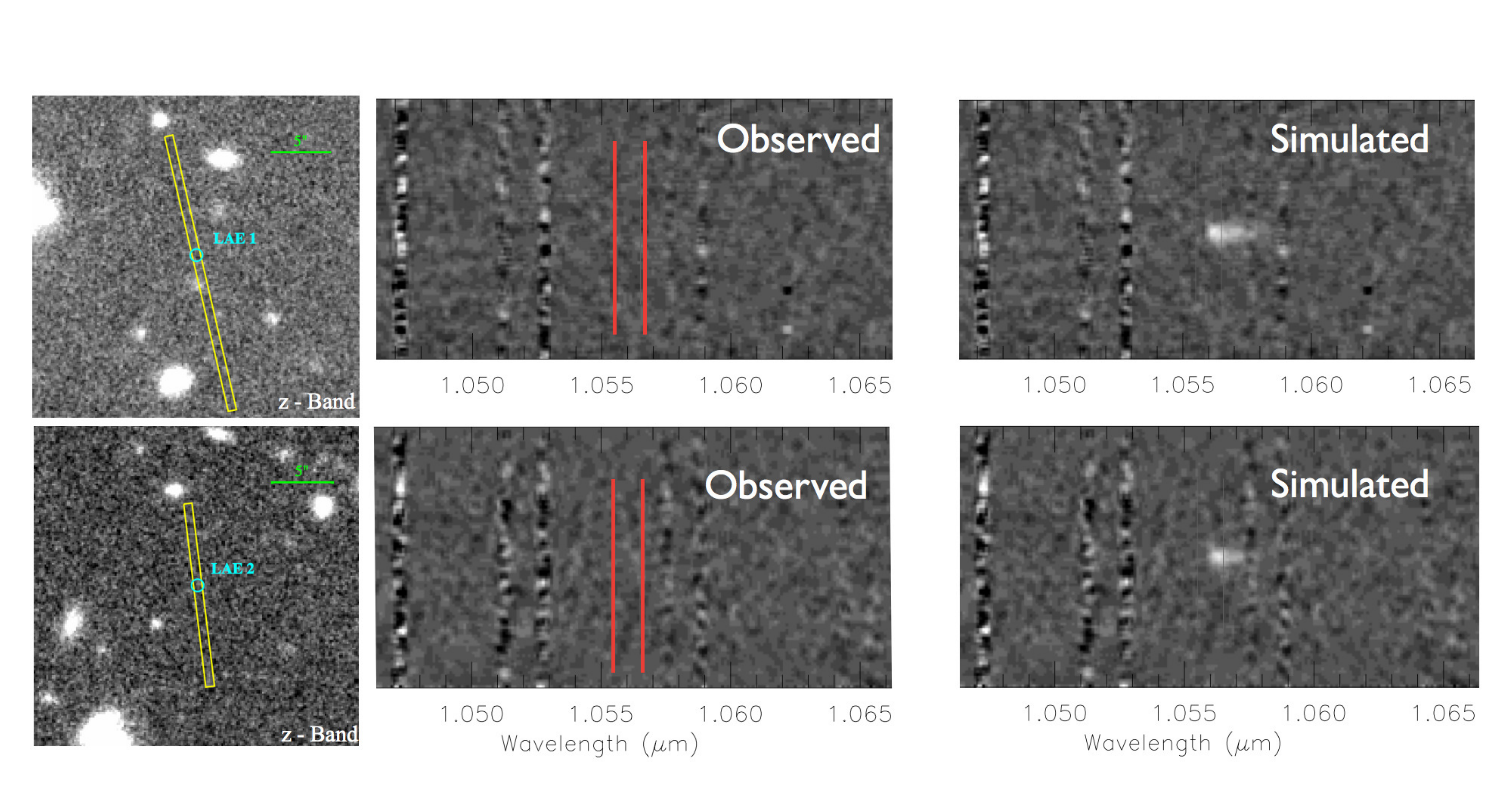}
\caption{SUBARU z$^{+}$-band images centered on LAE1 (top) and LAE2 (bottom) overlaid with the MOSFIRE slits configuration (left). Observed (center) and simulated 2D-spectra (right) are shown as well, both are binned to obtain R $\sim1500$. The wavelength range where the emission is expected from the UNB observations is marked with red lines. For the simulation shown here, we assumed a rest-frame FWHM of 1.5$\textup{\AA}$ for the Ly$\alpha$ line (represented as truncated gaussian) and a spatial extent of 1$\arcsec$. This simulation shows the clear detection of the line for both LAEs. \label{fig:Slit}}
\end{figure*}

\section{Observations \& analysis}

\subsection{Candidate selection by \citet{KRUG12}}
The two targets of our study are among the brightest LAE candidates at $z\sim7.7$ (12.1 and 8.6 $\times$ 10$^{-18}$ erg/s/cm$^{2}$, respectively, measured in UNB filters assuming negligible continuum). These targets were initially selected and published by \citet{KRUG12} and throughout this work we refer to these as LAE1 (brightest) and LAE2 (second brightest), respectively. Both LAEs were detected with an ultra narrow band (UNB) filter in the COSMOS field \citep{SCOVILLE07} using NEWFIRM \citep{AUTRY03}.  Details of the data reduction and selection are given in \citet{KRUG12}, but we give a brief summary of their results here.
The effective surface area of the UNB survey is $\sim$ 760 arcmin$^{2}$.  The UNB filter used for these candidates is centered at a wavelength of 1.056$\mu$m and has a width of $8-9\textup{\AA}$.  This is a dark region of the spectra between bright night sky lines, and selects objects with Ly$\alpha$ emission at a redshift $z\sim7.7$. The UNB data were acquired over a course of a year in three different sets of observations (February 2008, February \& March 2009). This means transient objects with periods of $<1$ year were rejected (see \citet{KRUG12} and later in this section). The total usable observations add up to $\sim$100 hours distributed over 32 nights, resulting in a limiting magnitude (defined as the 50\% completeness limit) of 22.4 AB in the UNB filter.
The area used to select these objects is covered by a second UNB filter centered at 1.063$\mu$m with the same width as well as deep ground based broad-band data  from Subaru in the optical (g, B, V, r, i, z) and UKIRT and Vista in the NIR (Y, J, H, K). This allows one to exclude continuum on the blue and red side of the potential Ly$\alpha$ emission line and should have eliminated low-z interlopers.
Both of the candidates are not detected in any of the broad band filters as well as the second UNB filter. This results in rest-frame equivalent width lower limits of $\sim7\textup{\AA}$ and $\sim5\textup{\AA}$ for LAE1 and LAE2, respectively.

\subsection{MOSFIRE observations \& data reduction}
We observed the two LAE candidates ($\alpha$ = 10$^{h}$00$^{m}$46$^{s}$.94, $\delta$ = +02$^{\circ}$08\arcmin48.84\arcsec and $\alpha$ = 10$^{h}$00$^{m}$20$^{s}$.52, $\delta$ = +02$^{\circ}$18\arcmin50.04\arcsec) with the MOSFIRE \citep{MCLEAN12} spectrograph on the Keck-I telescope on the nights of January 15 \& 16, 2013.  Each candidate was observed with a  separate mask created using the MOSFIRE Automatic GUI-based Mask Application (MAGMA\footnote{http://www2.keck.hawaii.edu/inst/mosfire/magma.html}, version 1.1) and aligned using bright 2MASS stars. The conditions were photometric on both nights, with an average seeing around $1.0\arcsec$. The observations were carried out in $Y$-band ($9710 - 11250\textup{\AA}$) using the YJ grating and a $0.7\arcsec$ slit width  resulting in a resolution $R\sim3270$.  We used 180s exposures with 16 Multiple Correlated Double Samples. The telescope was nodded by $\pm 1.25\arcsec$ with respect to the mask center position between exposures. The total integration times were $46 \times 180\textup{s} = 8280\textup{s} = 2.3\textup{h}$ for LAE1 and $40 \times 180\textup{s} = 7200\textup{s} = 2.0\textup{h}$ for LAE2, respectively.

Before creating the mask, we verified that the 2MASS, COSMOS, and NEWFIRM astrometric systems agreed to within measurable errors ($\sim 0.1\arcsec$). During the observations we make sure that the masks were properly aligned by using either alignment stars and/or bright filler targets. In addition several bright sources with known fluxes and morphologies from the zCOSMOS-bright spectroscopic survey \citep{LILLY09} were placed on the mask to verify slit losses (estimated to be 40-50\%). We observed 12 and 4 of these galaxies in the LAE1 and LAE2 masks, respectively. The comparison of the expected spatial position from MAGMA to the final spatial position on the reduced 2D spectra indicates that the alignment was better than $0.2\arcsec$ during the observations.

\begin{figure}
\includegraphics[scale=0.45,angle=270]{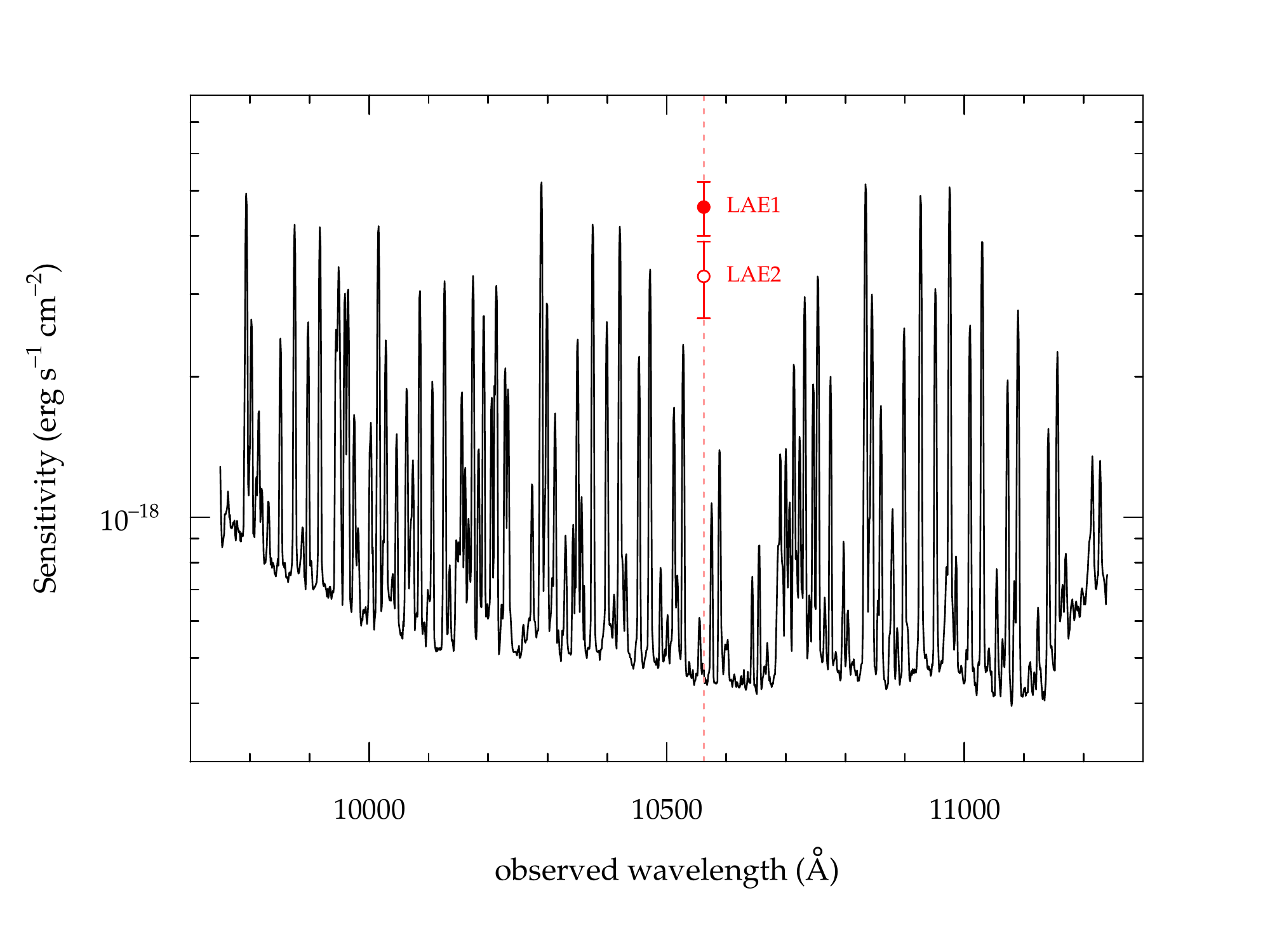}
\caption{The $Y$-band 1$\sigma$ sensitivity per 4.4$\textup{\AA}$ resolution element is shown. The measured sensitivity is consistent with that of the exposure time calculator corrected for slit losses. We should be able to detect the two LAE candidates at several $\sigma$ as shown by the red symbols representing their line fluxes as measured in the UNB filter by \citet{KRUG12}.\label{fig:Sensitivity}}
\end{figure}

We used the public MOSFIRE python data reduction pipeline\footnote{N. Konidaris, https://code.google.com/p/mosfire/} for sky subtraction, wavelength calibration, and co-addition of the single exposures. The pipeline performs an A--B / B--A subtraction and co-adds the single exposures using a sigma-clipped noise weighted mean after shifting them to a common pixel frame and masking bad pixels. The atmospheric OH sky lines are used for wavelength calibration. The final 2D spectra have a spatial resolution of 0.18$\arcsec$ per pixel and a spectral resolution of 1.09 $\textup{\AA}$ per pixel. Figure \ref{fig:Slit} shows the final 2D spectra (degraded to $R\sim1500$) together with the slit positions on sky.  We measured an RMS noise of $5-10\times10^{-19}$erg/s/cm$^{2}$ (4.4$\textup{\AA}$ resolution element) in the 10545-10565$\textup{\AA}$ wavelength region, in good agreement with the estimated noise from the the MOSFIRE exposure time calculator\footnote{ETC version 2.3 by G. C. Rudie, http://www2.keck.hawaii.edu/inst/mosfire/etc.html}, corrected for our estimated slit losses. 
Absolute flux was measured using the white dwarf spectrophotometric standard star GD71. The standard star was observed during the same nights with identical settings and reduced in the same way as the science exposures. We present the sensitivity curve for the MOSFIRE $Y$-band in Figure \ref{fig:Sensitivity} together with the line fluxes of the two targets derived from UNB filters. This shows that we would have clearly detected the two LAEs as it is further discussed below.

\subsection{Tests and Simulations: Establishing our detection limits}
Assuming the observed fluxes given in \citet{KRUG12} at 10560$\textup{\AA}$ and based on our measured noise and our seeing of 1$\arcsec$, we expect to detect the two sources at a signal-to-noise of 12.4 and 8.2, respectively, with a line width of 200km/s (e.g., \citet{HU10}). Even with a seeing as bad as 2$\arcsec$, the expected signal-to-noise is still 8.8 and 5.8, respectively.
To verify the SNR calculation and lack of detection, we simulated the expected 2D spectra by adding lines to the reduced 2D spectra.
For these simulations we assumed the total measured flux was distributed over a truncated gaussian with a rest-frame FWHM ranging from $0.5 - 3.0\textup{\AA}$ (observed from stacked spectra it is $\sim 1.5\textup{\AA}$, e.g., \citet{HU10}). For the spatial extent we assume a gaussian with FWHM of 1$\arcsec$ corresponding to our seeing. The following results of our simulation are not sensitive to the actual spatial extent. We find that the total flux of such a line would have to be less than $\sim2 - 4\times10^{-18}$~erg/s/cm$^{2}$ for not to be visible in our data (for the range in rest-frame line FWHM). Vice versa, to miss LAE1 (LAE2) in our data, we would require a rest-frame FWHM of more than 10$\textup{\AA}$ (7$\textup{\AA}$).
Figure \ref{fig:Slit} shows the simulated spectra rescaled to $R=1500$ assuming the line fluxes measured by \citet{KRUG12}, a line rest-frame FWHM of 1.5$\textup{\AA}$, and spatial extent of 1$\arcsec$. This shows a clear expected detection of both Ly-$\alpha$ lines.

\subsection{No detection of Ly$\alpha$ in LAE1 and LAE2}
Our firm non-detection of line emission in the targeted LAEs yields an upper limit in Ly$\alpha$ line flux of $2 - 4\times10^{-18}$~erg/s/cm$^{2}$.
We therefore rule these candidates out on a 7 and 5$\sigma$ level, respectively.
This puts on solid ground a recent less significant non-detection by \citet{JIANG13} in 7.5 hours of LBT observation with the LUCI NIR spectrograph.
Given these limits, the sources must either be a transient event with decay times of $>1$ year, very short periodic ($\ll1$ year) with a large change in flux, or artifacts and/or noise spikes in the data.
Considering transients, the most likely events with similar rates are super-luminous Supernovae (SSNe) or AGNs.
Low redshift SNe are favorable because the rest-frame NIR emission is decaying less rapidly than the optical \citep{TANAKA12}. These events can account for the magnitude change measured in the UNB filters \citep{QUIMBY07,GEZARI09,MILLER09}. However, a simple calculation suggests that a $z\sim0.3$ SSN is visible for maximum $\sim230$ days (observed) including the rise of luminosity before its peak. However, \citet{KRUG12} searched for objects with variability on these time scales and removed them, therefore we believe these are an unlikely source of contamination, although up to three such events could have happened within 0.2 deg$^{2}$ during the year of observations depending on IMF \citep{TANAKA12}.
Furthermore, short period AGNs can be excluded as a source of contamination, because of the amplitude of the variability which exceeds that in known AGNs \citep{VANDENBERK04,WILHITE08,BAUER09}.
We thus conclude that the detections are most likely artifacts and noise. There are several reasons why this could happen. First of all, detections near the edges of an image can be caused by enhanced noise. Also, estimates of the limiting magnitude by using 50\% completeness simulations and/or the use of inappropriate aperture sizes with respect to the seeing may lead to false detections.
In the case of the \citet{KRUG12} candidates, the authors use 50\% completeness simulations to estimate their limiting magnitudes. Also, their candidates seem to lie systematically close ($\sim3$ arcmin) to the chip gaps between the four NEWFIRM arrays.
Combined with the findings of \citet{CLEMENT12} and \citet{JIANG13}, who also find no real detections, this raises significant questions about the reliability of the narrow band filter technique with NIR detectors for detecting LAEs at $z>7$. Note that for $z<7$, where large spectroscopic follow-up studies of LAE candidates are possible, the fraction of low-z interlopers and spurious objects is usually $<40$\%.

Whatever the reason, the non-detection of LAE1 and LAE2 in the MOSFIRE spectra places important limits on the LAE LF and implies strong evolution of it at $z>6$ as it will be discussed in the following section.

\begin{figure*}
\includegraphics[scale=0.6,angle=270]{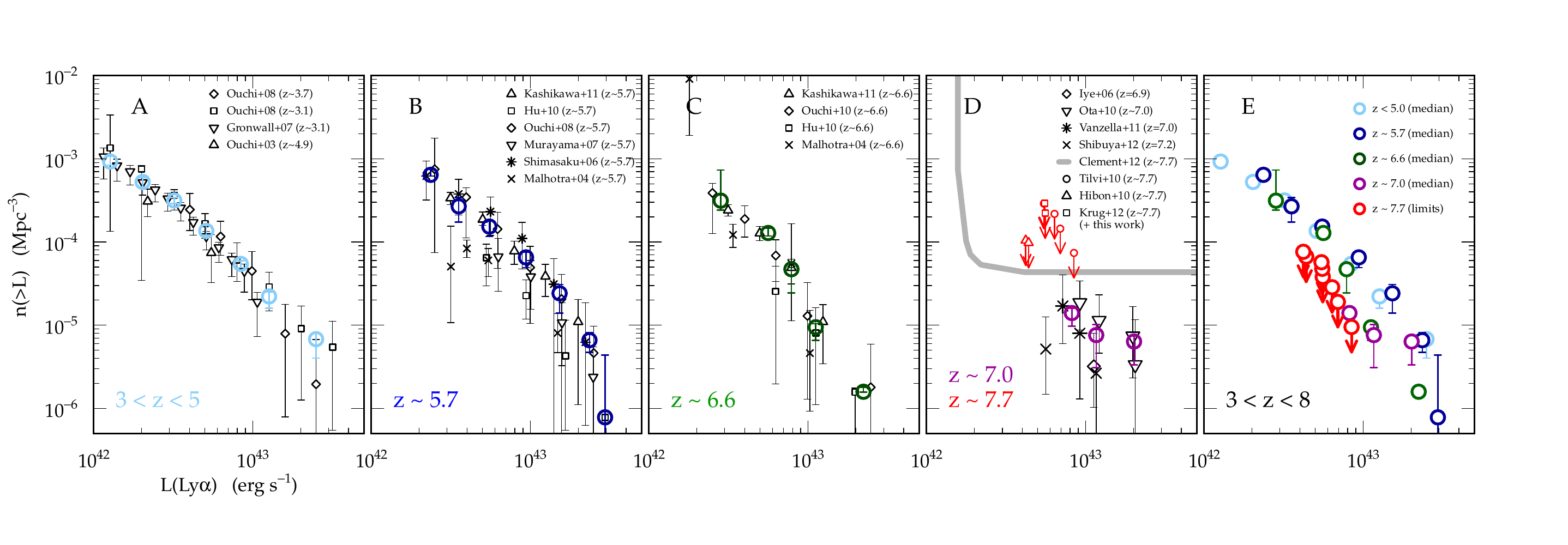}
\caption{Compilation of different studies measuring the Ly$\alpha$ LFs at $z~\sim~$3.1, 3.7, 4.9, 5.7, 6.6, 7.0,  and 7.7 (panels A through D). Black symbols denote single studies whereas colored symbols represent their weighted medians. The error bars on the colored symbols show the standard deviation on the median. The red symbols in panel D represent limits from single \textit{candidates} at $z\sim7.7$ from three different studies (see legend). These limits are combined and shown as red circles in panel E together with the median measurements at $z<7.7$ from the other panels. The new limits at $z\sim7.7$ are consistent with an evolution of the bright end of the LAE LF at $z>6$.\label{fig:literature}}
\end{figure*}

\section{The evolution of the Ly$\alpha$ LF from $z=3.1$ to $z=7.7$}

A large number of studies have looked at the Ly-$\alpha$ luminosity function at $z<7$.
A summary of the surveys at $z\sim3-5$ is given in Table \ref{tab:LF1} and the mean data points adopted for this redshift range are shown in Figure \ref{fig:literature}, panel A.
It can be seen that the LAE LF changes only slightly in this redshift range. Schechter functions fitted to the data as a function of redshift result in less than 15\% change in $L^{*}$ and $\phi^{*}$, respectively \citep{OUCHI08}.
Note that in this and the following comparisons of LFs, we account for eventual differences in the cosmologies assumed by the authors.
Furthermore, some authors apply a correction to their Ly$\alpha$ luminosities to account for absorption of the Ly$\alpha$ forest. This correction is debated as it is shown recently that the Ly$\alpha$ line profile is asymmetric at $z\sim0$ where IGM absorption is negligible. This suggests that Ly$\alpha$ is already redshifted when escaping the galaxy and most probably make the above correction factor superfluous and result in an overestimation of the LAE luminosity \citep{SCARLATA14}. The LFs presented in this paper are not corrected by this factor.

At $5<z<7$ there are severals major studies \citep{TANIGUCHI05, SHIMASAKU06,MURAYAMA07,OUCHI08,HU10,OUCHI10,KASHIKAWA11}. All of these use the Subaru/Suprime-Cam camera with the NB812/NB921 filters. Additional constraints come from \citet{MALHOTRA04} compiling a large sample of LAE surveys. The above studies are summarized in Table \ref{tab:LF2}.
These large studies have significant disagreements in the derived luminosity functions with the various studies citing contamination rates, selection functions, and spectroscopic incompleteness as possible sources of disagreement. The \citet{HU10} study uses several widely spaced fields to rule out cosmic variance as the source of the discrepancy.
In Figure \ref{fig:literature} we combine the various studies, and find that while the fits to the LAE LFs done by the different authors disagree, the data are consistent within errors, indicating counting statistics and fitting methods are the likely source of the discrepancy. We adopt weighted averages of the data points for the redshifts $3.1<z<4.9$, $z\sim5.7$, and $z\sim6.6$ as indicated by the colored symbols in Figure \ref{fig:literature} panels A through C. In panel D we also show LAE detections by \citet{IYE06}, \citet{OTA10}, \citet{VANZELLA11}\footnote{We note that these two galaxies are not selected by a systematic NB search. However, they could be detected by these according to their properties (Ly$\alpha$ fluxes, broad-band magnitudes, and EWs) and we therefore include them here.}, and \citet{SHIBUYA11} at $z\sim7$ with their weighted averages.
We note that there are differences in the normalization of the above studies which are likely linked to sample incompleteness and contamination (estimated to be less than 20\% for these studies). These uncertainties are captured in the individual error bars, which we take into account in the final error bars of the weighted averages.
At $z\sim7.7$, we combine the candidate detections from \citet{HIBON10} and \citet{TILVI10} with the two remaining candidates from \citet{KRUG12} by adding up the comoving volumes of the studies. The new limits at $z\sim7.7$ are shown in Figure \ref{fig:literature} panel E. The single points are shown in Figure \ref{fig:literature} panel D together with the limit from \citet{CLEMENT12} shown as gray line.
Finally in Figure \ref{fig:literature} panel E, we show our combined luminosity functions over the redshift range $3.1 < z < 7.7$.

This clearly shows a rapid evolution in the number density of bright LAEs at $6 < z < 8$.  However, it is unclear whether this evolution is driven by changes in the IGM opacity, or evolution in the density of the underlying galaxy population. We will disentangle these two effects in the following section.

\section{The fraction of neutral hydrogen at $z\sim8$}

Ly$\alpha$ emission is produced in young galaxies with a substantial amount of on-going star-formation. It is therefore the amount of UV radiation and the ISM of a galaxy which constrains the amount of Ly$\alpha$ emission.
As the Ly$\alpha$ photons escape from the galaxy, they get scattered in areas of dense neutral hydrogen in the IGM. The amount of neutral hydrogen around galaxies sets the amount of Ly$\alpha$ emission that can be measured by our telescopes.
As soon as galaxies are formed, they start to re-ionize larger and larger bubbles of neutral hydrogen around themselves and the transparency for Ly$\alpha$ photons is increased. 
By recording the amount of Ly$\alpha$ emission, i.e., the rest-frame equivalent width (EW$_{0}$) distribution,  as a function of redshift, it is therefore possible to estimate the change in the volume fraction of neutral hydrogen, $x_{HI}$, and therefore map the re-ionization process.
 
However, the change in the fraction of Ly$\alpha$ emitting galaxies also depends on the density of the underlying galaxy population as well as on internal (ISM) properties of the galaxies, like star formation rate and dust content.
Studies of the Ly$\alpha$ emission properties of \textit{UV-continuum selected LBGs} suggest that the Ly$\alpha$ emission is rising with redshift in galaxies at $z=4-6$ \citep{STARK10,MALLERY12,SCHENKER12}, where the universe is thought to be fully re-ionized.  In particular, \citet{ZHENG13} note that the EW distribution in this redshift range ($4 < z < 6$) is skewed to larger rest-frame EW values for higher redshifts.
This suggests evolution of the internal properties of galaxies (e.g., dust, \citet{BOUWENS12b,FINKELSTEIN12,MALLERY12}) enhancing the amount of Ly$\alpha$ emission with increasing redshift (e.g., \citet{TREU12}).

In order to constrain the fraction of neutral hydrogen at $z\sim8$, we have to separate these effects from the IGM.
We therefore first model the intrinsic (i.e., without IGM absorption) Ly$\alpha$ LF ($\S$4.1). Later, we will compare this intrinsic LF to the observed LFs at different redshifts ($\S4.2$) and, combined with two possible implementations of the re-ionization process, constrain $x_{HI}$ ($\S4.3$).

\begin{figure*}
\includegraphics[scale=0.6,angle=270]{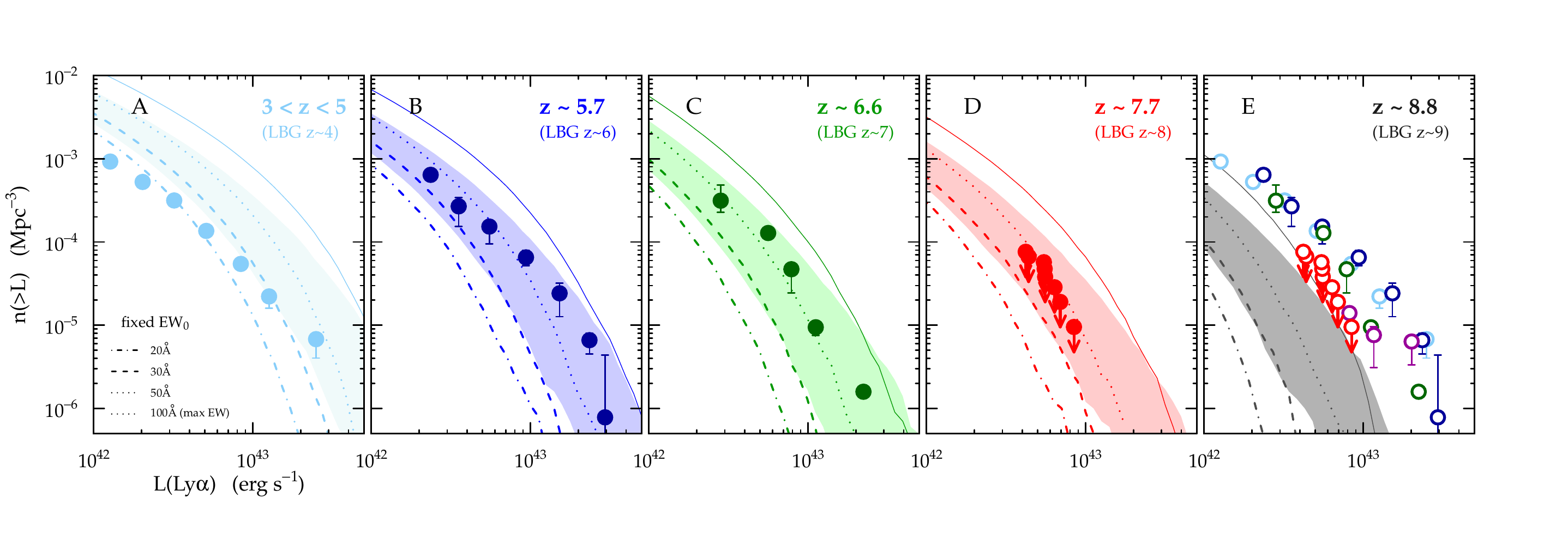}
\caption{Comparison of the measured Ly$\alpha$ LFs (symbols) to our empirical model combining the UV continuum LF with the observed rest-frame equivalent-width distribution and assuming $X_{Ly\alpha}=1$ (see text for more details). The range of LFs due to two different EW distributions from literature is indicated by the shaded regions. The (intrinsic) EW distributions are the same for all redshifts in our model. A constant EW$_{0}$ of 20, 30, and 50$\textup{\AA}$ is shown as dot-dashed, dashed, and dotted line, respectively. The solid line denotes a fixed EW of $100\textup{\AA}$ corresponding to the maximal EW with a Salpeter IMF ($M_{upper}=120$M$_{\odot}$) and $Z=1/20$ solar metallicity \citep{MALHOTRA02}. This comparison shows, that the LAE LF is correlated with the Ly$\alpha$ LF derived by UV selected galaxies. We use this fact to extrapolate the LAE LF to $z\sim8.8$ as it is shown in panel E and predict an upper limit for the number of expected LAEs in different planned surveys. There is, however, a second order effect: The observed LAE LF is slightly changing with respect to the model. This can be interpreted as changing properties of the IGM acting on the rest-frame equivalent width distribution of the galaxies. This can be used to estimate the neutral hydrogen fraction of the IGM as it is outlined further in the text and Figure~\ref{fig:reionization}.
\label{fig:model}}
\end{figure*}

\subsection{A model of the LAE galaxy population}

To separate ISM from IGM effects on the Ly$\alpha$ LF (see also \citet{DIJKSTRA12}), we first create an empirical model of the LAE LF based on the UV LF and the Ly$\alpha$ rest-frame equivalent width (EW$_{0}$) distribution at $z<6$, where the IGM is fully re-ionized.
In brief, we assume the $z=4-9$ UV-continuum LFs of LBGs derived by \citet{BOUWENS07,BOUWENS11} and \citet{OESCH13}. These LFs can be well explained by assuming that the luminosity and stellar mass of a galaxy is directly related to its dark-matter halo assembly and gas infall rate \citep{TACCHELLA13}. Especially the LF at $z>7$ are therefore put on more solid ground. We then convolve these UV LFs with two observed Ly$\alpha$ EW$_{0}$ distributions of \citet{MALLERY12} ($4 < z < 6$) and \citet{STARK10} ($3 < z < 7$) by using a Monte Carlo sampling method to estimate a LAE LF.

We first draw random galaxies from the UV-continuum LFs. The number of galaxies is defined by the integral of the UV luminosity function at the different redshifts.
On the bright end we integrate to $M_{UV}=-30.0$, above which the contribution of galaxies becomes negligible.
On the faint end, we set the integration limit to $M_{UV}=-15.0$. We note, that this is $\sim$2 magnitudes below the Ly$\alpha$ luminosity which is observed at all redshifts. Changing $M_{UV}$ above this limit does not change the output of our model. This faint $M_{UV}$ limit however means extrapolating the observed UV-continuum LFs used from the literature (usually going down to $M_{UV}=-18.0$). So we also verified that the implications of our model are insensitive to changes of the faint end slopes of the UV-continuum LFs and other LF parameters between different studies \citep{BRADLEY12,MCLURE13, SCHENKER12}.
For each of the galaxies drawn from the UV-continuum LF we then pick a random rest-frame equivalent width from the input distributions and compute the cumulative Ly$\alpha$ LFs.
We assume no correlation between EW$_{0}$ and UV-luminosity for simplicity, although there are hints of less luminous galaxies reaching larger EW$_{0}$ compare to more luminous ones (\citet{SCHAERER11} but see \cite{NILSSON09} and \cite{ZHENG13} for a contradictory study).
We also assume that every galaxy is emitting Ly$\alpha$ (which is then absorbed in the IGM and the EW distribution captures the ISM physics), i.e., the fraction of Ly$\alpha$ emission ($X_{Ly\alpha}$) is 100\% for our model.

Our models are shown as shaded regions in Figure \ref{fig:model}, panels A through E.
The points show the same weighted averages as in Figure \ref{fig:literature} and we find that our model is very sensitive to the assumed EW$_{0}$ distribution. This is illustrated by the broad swath of the shaded region indicating the range of values obtained by the \citet{MALLERY12} and \citet{STARK10} EW$_{0}$ distributions. This is not surprising, as from the comparison of the two EW$_{0}$ distributions it can be seen that \citet{MALLERY12} is missing high EW$_{0}$ compared to \citet{STARK10} which results in a much lower Ly$\alpha$ LF estimate.
In the following we will assume the \citet{STARK10} EW$_{0}$ distribution as basis because it samples fainter galaxies which contribute to the majority of objects in our sample while \citet{MALLERY12} is restricted to UV continuum redshifts and therefore brighter galaxies.
To illustrate the dependence on EW$_{0}$ further, the dotted, dashed, and dash-dotted lines in Figure \ref{fig:model} show \textit{constant} input rest-frame equivalent widths with EW$_{0} = 20, 30, 50\textup{\AA}$. 

\begin{figure*}
\includegraphics[scale=0.68,angle=270]{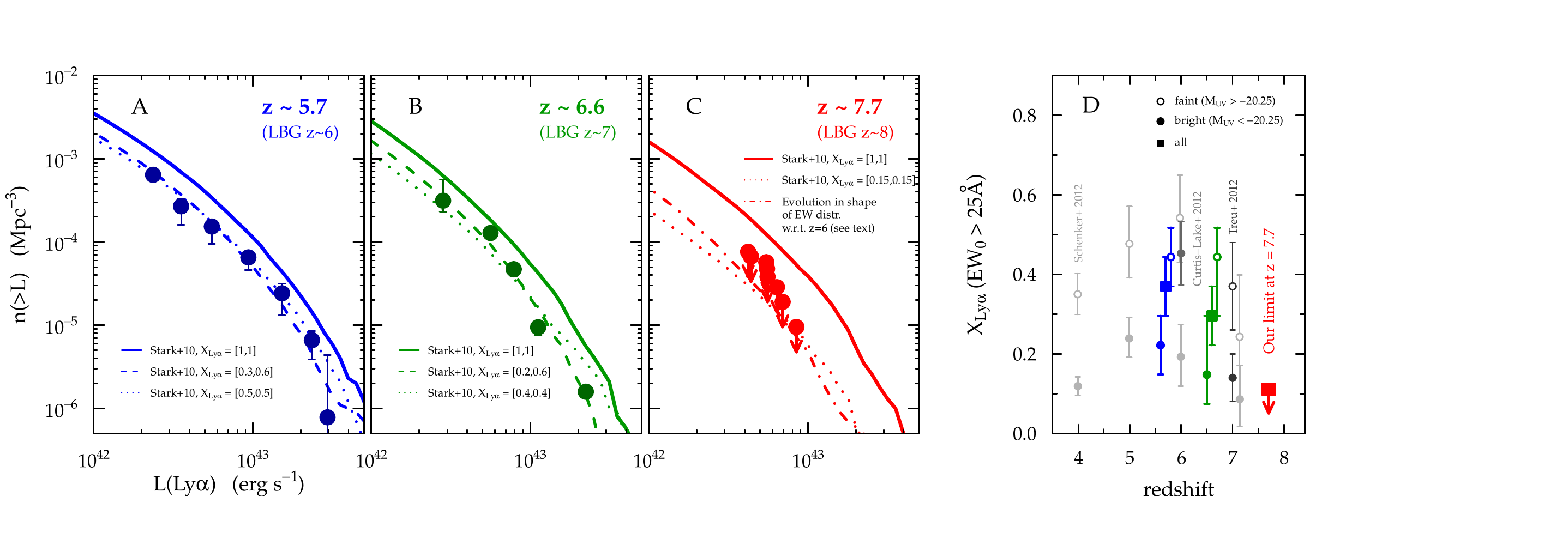}
\caption{Two methods to constrain the change in the EW distribution of Ly$\alpha$ emitting galaxies with redshift by comparing our model (solid, \citet{STARK10} EW$_{0}$ distribution as basis, 100\% Ly$\alpha$ emission) to the observe LAE LF (symbols).
\textit{(i)} The dashed and dotted lines show our model tuned to fit the data by adjusting $X_{Ly\alpha}$ (dotted: overall, dashed: split in bright and faint magnitude, see text). Panel D summarizes its evolution as a function of redshift from our work (colored symbols) compared to observations by \citet{SCHENKER11} (light gray), \citet{CURTISLAKE12} (dark gray) and \citet{TREU12} (black) for galaxies with EW$_{0}>25\textup{\AA}$. A drop in the fraction of Ly$\alpha$ emitting galaxies of a factor 4 above $z=6$ is clearly visible.
\textit{(ii)} The dot-dashed line in panel C shows our model tuned to fit the data by skewing the EW$_{0}$ distribution to lower EW$_{0}$ values (i.e., adjusting its width). Both methods of modifying the EW distribution result in consistent estimates of the lower limit of neutral hydrogen fraction at $z\sim7.7$ of 50-70\% (see text).\label{fig:reionization}}
\end{figure*}

\subsection{Interpreting the evolution of LAEs}

We find good overall agreement between our ``predicted'' LAE LF and the observed values up to $z\sim7$. But note in Figure \ref{fig:model} the observed LAE LF moves from the bottom of the predicted range at $3<z<5$ to the top at $z\sim5.7$.  This indicates the EW$_{0}$ distribution appears to be skewing to higher values as found by \citet{ZHENG13} (compare with the lines at constant EW$_{0}$ in Figure \ref{fig:model}) and is likely caused by decreasing amounts of dust.
In contrast, at $z>7$ the LAE LF appears to return to the middle or bottom range of the shaded region predicted by our model. Assuming the (intrinsic) EW distribution does not change, then a change in the IGM is needed to reproduce the observation.
This indicates the IGM is becoming more opaque at $z>7$, suggesting re-ionization finished at $z\sim6-7$.

\subsection{Constraint on $x_{HI}$ and Ly$\alpha$ optical depth at $z\sim7.7$}
Turning to a more qualitative analysis, we use our model to constrain the change in neutral hydrogen fraction in the IGM at $z>6$.

For this, we consider two different possibilities of how we think Ly$\alpha$ photons get absorbed in the IGM. The two different approaches lead to different imprints of re-ionization in the Ly$\alpha$ luminosity functions.
We consider \textit{(i)}  a ``black and white'' process where Ly$\alpha$ emission of a galaxy is either absorbed or not (``patchy/absorption model'') and \textit{(ii)} a smooth process where the Ly$\alpha$ emission is attenuated by a certain degree (``smooth/attenuation model'').

The former process will decrease the number of Ly$\alpha$ emitting galaxies irrespective of their emission strength. It will lead to a ``global'' shift of the LAE LF.
The later process will lower the Ly$\alpha$ emission in all of the galaxies, preferentially removing galaxies with high Ly$\alpha$ rest-frame equivalent width. It will lead to a change in normalization \textit{and} shape of the LAE LF.

For both of these models we can constrain $x_{HI}$ independently.
We estimate $x_{HI}$ for the former by using the simulations by \citet{MCQUINN07}, for the later we apply the models by \citet{DIJKSTRA11}.

We note that with the current data it is not possible to (dis)prove one or the other approach.
But we will see that both approaches will lead to the consistent results.

\subsubsection{A patchy model of re-ionization}

In this case Ly$\alpha$ is blocked by the neutral IGM which results in a decrease of the Ly$\alpha$ LF for all luminosities.
We tune our model LF to fit the observed LAE LFs at $z\sim$ 5.7, 6.6, and 7.7 by adjusting $X_{Ly\alpha}$ (the total fraction of galaxies for which Ly$\alpha$ is not absorbed), which is (at first) independent of magnitude (see Figure \ref{fig:reionization} panels A through C, dotted curves). We find that $X_{Ly\alpha}$ is almost undistinguishable between $5.7 < z < 6.6$ but drops by a factor of 4 beyond $z=7$ as it is shown in Figure \ref{fig:reionization}, panel D by the filled squares.
Furthermore, we follow the approach of \citet{SCHENKER11} and introduce two different values $X_{Ly\alpha}^{bright}$ and $X_{Ly\alpha}^{faint}$ for simulated galaxies with $M_{UV} < -20.25$AB and $M_{UV} > -20.25$AB in order to compare the fraction of Ly$\alpha$ emitters from our empirical model to real observations at $z<7$. This is shown in Figure \ref{fig:reionization} panels A and B by the dashed line (we do not apply this split at $z\sim7.7$ because of the sparse data). The values for $X_{Ly\alpha}^{bright}$ and $X_{Ly\alpha}^{faint}$ for EW$_{0}>25\textup{\AA}$ are shown in panel D (filled and open circles, respectively). The error bars are estimated by changing the M$_{UV}$ cut in a range of $M_{UV}=-20.25 \pm 2$.
Also shown are the observations by \citet{SCHENKER11} (light gray), \citet{TREU12} (black), and \citet{CURTISLAKE12} (dark gray) for galaxies with EW$_{0}>25\textup{\AA}$ and the same magnitude cut.

In general, we find a good agreement of $X_{Ly\alpha}(z)$ with the values observed in UV-continuum selected LBGs at $z<7$. We find a significant drop of a factor of $4\pm1$ in the fraction of Ly$\alpha$ emitters at $z\sim7.7$ compared to $z=6$.
Note that the \citet{CURTISLAKE12} estimate of $X_{Ly\alpha}$ for bright galaxies is a factor of $\sim2$ higher than the estimates from the other studies. Different selection and sample variance are a very likely cause for this discrepancy.
Nonetheless, their results support a strong drop of $X_{Ly\alpha}$ above $z=7$.

This change in LF can be converted into a neutral hydrogen fraction ($x_{HI}$) by using the results from 186-Mpc radiative transfer simulations by \citet{MCQUINN07} as follows: their figure 4 shows the relative change of the Ly$\alpha$ LF as a function of neutral hydrogen fraction at $z=6.6$ assuming full re-ionization at $z=6$. For example $x_{HI}=$ 0.18, 0.38, 0.53, 0.67, and 0.80 result in a re-scaling of the LF with factors of 0.76, 0.50, 0.33, 0.20, and 0.05, respectively. We then assume that this re-scaling of the LF is directly proportional to the change in the fraction of Ly$\alpha$ emitters, i.e., $X_{Ly\alpha,z=7.7}/X_{Ly\alpha,z=6} \sim4$ (see Figure \ref{fig:reionization}, panel D, blue and red squares).
Assuming the dust extinction properties at $z\sim7.7$ are the same as at $z=6$, we conclude that the drop in $X_{Ly\alpha}$ implies a neutral hydrogen fraction of \textit{at least} $x_{HI}=0.60\pm0.07$ at $z\sim7.7$. Assuming the dust content of galaxies above $z=6$ is further decreasing and therefore extrapolating $X_{Ly\alpha}(z)$ from the values at $4 < z < 6$ (see \citet{STARK10}) implies even higher limits ($x_{HI}=0.71\pm0.04$).
Note that the small change in $X_{Ly\alpha}$ between $z\sim5.7$ and $z\sim6.6$ is indicative of little neutral hydrogen. 
This is in line with the results by \citet{MCQUINN07} who suggest the universe is fully ionized at these redshifts.

Note that we can estimate $x_{HI}$ without applying our model, by directly taking the ratio of the LAE LFs at $z\sim5.7$ and $\sim7.7$ and applying again the simulations by \citep{MCQUINN07}. This approach leads to consistent results.

\begin{figure}
\includegraphics[scale=0.68,angle=270]{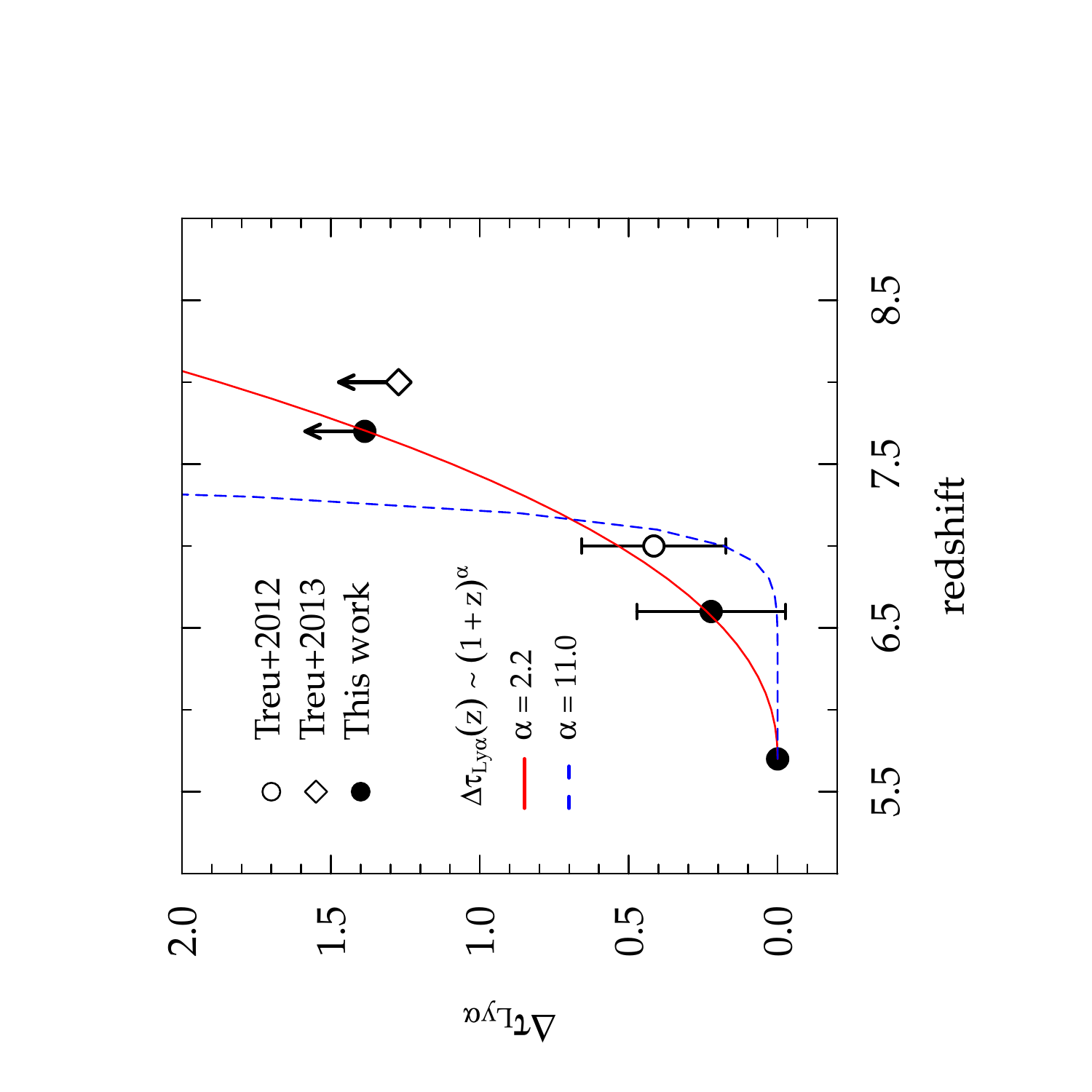}
\caption{Change in Ly$\alpha$ optical depth with redshift with respect to $z=6$ assuming the universe is fully re-ionized by then. Under this assumption, we use the \citet{TREU12} formalism to find the mean change in Ly$\alpha$ optical depth with respect to $z=6$ which we assume to be proportional to the change in fraction of Ly$\alpha$ emitting galaxies. Our limit at $z\sim8$ is important to constraint $\tau_{Ly\alpha}(z)$ which we find to be best fit as $(1+z)^{\alpha}$, $\alpha=2.2\pm0.5$ (solid red line). The strong evolution of at least 1.3 beyond $z=6$ is apparent and could be indicative of a dramatic change in the properties of the IGM. Shown along with our best fit is the exponent from the best fit to the evolution of the Gunn-Peterson optical depth measured on Ly$\alpha$, Ly$\beta$, and Ly$\gamma$ transitions in quasars \citep{GOTO11,FAN06}.\label{fig:tau}}
\end{figure}

Having established a lower limit on $x_{HI}$, we can use the patchy model further to constrain the Ly$\alpha$ optical depth.
Assuming the change in $X_{Ly\alpha}$ above $z=6$ is due to the IGM, it can be associated to the \textit{average change of Ly$\alpha$ optical depth $\left<e^{-\Delta\tau_{Ly\alpha}}\right>$ under the assumption that re-ionization is completed at $z\sim6$} (i.e., $\tau_{Ly\alpha,z=5.7} = 0$ and $\Delta\tau_{Ly\alpha}(z) = \tau_{Ly\alpha}(z)-\tau_{Ly\alpha,z=5.7}$). Note, that this approach is identical to \citet{TREU12} and we can set $X_{Ly\alpha}(z)/X_{Ly\alpha,z=5.7} \equiv \epsilon_{p}(z)$, where $\epsilon_{p}$ is defined as in \citet{TREU12} and $\epsilon_{p,z=6}=1$ by construction.
From Figure \ref{fig:reionization}, panel D we find $\epsilon_{p}$ = 0.8$\pm$0.2 for $z\sim6.6$ (blue and green squares) and $\epsilon_{p}$ = 0.25$\pm$0.05 for $z\sim7.7$ (blue and red squares), respectively.
Our $z\sim6.6$ ($z\sim7.7$) value is consistent with the $z\sim7$ ($z\sim8$) value of 0.66$\pm0.16$ ($<0.28$) found by \citet{TREU12} \citep{TREU13} within errors.
We then compute the Ly$\alpha$ optical depth by equating $\epsilon_{p}(z) = \left<e^{-\Delta\tau_{Ly\alpha}(z)}\right>$.
The final result of $\Delta\tau_{Ly\alpha}(z)$ w.r.t. $z\sim6$ is shown in Figure \ref{fig:tau}.
Our limit at $z\sim7.7$ is important to constraint $\Delta\tau_{Ly\alpha}(z)$ as the values at $z\sim6$ and 7 are almost indistinguishable. The overall change in optical depth as a function of redshift can be expressed by $\Delta\tau_{Ly\alpha}(z) \propto (1+z)^{\alpha}$ with $\alpha=2.2\pm0.5$. Note that this exponent is a lower limit because of the upper limit in the LAE LF at $z\sim7.7$.
We find an increase in optical depth of at least 1.3 between $z=6$ and $z\sim8$.
Our best fit model is fully consistent with the Gunn-Peterson optical depth measurements in quasars \citep{GOTO11,FAN06}, however the functional forms of the estimates lead to different exponents (see Figure \ref{fig:tau}).

\subsubsection{A smooth model of re-ionization}

In this case there is no global scaling of the LF as before, however a steepening of the LF may occur because the EW$_{0}$ distribution gets skewed to lower EW$_{0}$ as the redshift increases beyond $z = 6$ (see also \citet{ZHENG13}). We represent the Stark et al. EW distribution in the same manner as \citet{TREU12} by using a gaussian truncated at negative values. In contrast to the case outlined before, we now change the width of the EW$_{0}$ distribution (similar to the ``smooth model'' in \citet{TREU12}). As in the case above, we have to take the difference in evolution between $z=6$ and $z=7.7$ (assuming the IGM is fully re-ionized at $z=6$). We therefore start directly with the $z=6$ EW$_{0}$ distribution (see Figure \ref{fig:reionization}, panel A, dotted curve) and tune it to fit the $z\sim7.7$ limits by changing its width (dashed-dotted line in Figure \ref{fig:reionization}, panel C). From the final EW$_{0}$ distribution at $z\sim7.7$, we compute the cumulative fraction $P(>EW_{0})$ which has now changed w.r.t. $z=6$ as we have adjusted the width of the EW$_{0}$ distribution. This fractions can be converted into $x_{HI}$ by using the models by \citet{DIJKSTRA11} (using semi-numerical simulations by \citet{MESINGER11}) combining galactic outflow models and large-scale semi-numeric simulations of reionization. From our final EW$_{0}$ distribution fitting the limits at $z\sim7.7$ we find $P(>100\textup{\AA})=0.02\pm0.01$, $P(>75\textup{\AA})=0.07\pm0.02$, and $P(>50\textup{\AA})=0.20\pm0.05$ which translates, by adopting figure 5 in \citet{PENTERICCI11}, into upper limit neutral hydrogen fractions of $x_{HI}$ =  $0.7\pm0.1$, $0.6\pm0.1$, and $0.5\pm0.2$, respectively. Note that $x_{HI}$ is more difficult to estimate for smaller EW$_{0}$ cuts as $P(>EW_{0})$ approaches unity for all $x_{HI}$ by construction \citep{PENTERICCI11}. Taking this into account, the limits we find with our second approach are consistent with the results above.

\subsubsection{Summary of our findings}

In summary, we have looked at two different ways how re-ionization can be imprinted in the change of Ly$\alpha$ LF. We have considered an absorption model resulting in a global shift of the Ly$\alpha$ LF and an attenuation model resulting in a skewing of the EW$_{0}$ distribution and there for a steepening of the Ly$\alpha$ LF.
Note, that both approaches can fit the observed LAE LFs within its uncertainty and we are not able to judge which of the models is right.
However, a skewing of the EW distribution is likely as it seems from the observational data at $z\sim5.7$ and $z\sim6.6$ that the evolution of the bright end is stronger than at the faint end of the LAE LF.
In either way, we are able to constrain $x_{HI}$ using both approaches, resulting in lower limits for the neutral hydrogen fraction between $x_{HI} = 0.53$ and $x_{HI} = 0.70$ at $z\sim7.7$.

Finally, we stress that our results are based on the assumption that all changes in $X_{Ly\alpha}$ and the EW$_{0}$ distribution are caused by a change in the ionization state of the IGM at $z>6$.
However, and alternative explanation involves an increase of the escape fraction of ionizing photons and would lead to a drop in $X_{Ly\alpha}$ and thus an overestimation of $x_{HI}$ \citep{DIJKSTRA14}. 
Without a changing ionization state of the IGM the escape fraction needed to explain the observations is at odds with other studies \citep{WYITHE10,KUHLEN12,ROBERTSON13,DIJKSTRA14}.
However, a mixture of changing $x_{HI}$ ($\sim 0.2$) and $f_{esc}$ ($\sim 0.2-0.3$) would be consistent with our results and direct escape fraction measurements.


\section{Expected number detections of LAEs at $z\sim8.8$ in other surveys}

Given these results at $z\sim7.7$ it is important to push to higher redshifts to better constrain the evolution of the LAE LF. Assuming that the LAE LF continues to trace the LBG LF at $z > 8$, we can put upper limits on the number of LAEs that should be found in planned surveys.
The final UltraVISTA NB118 survey \citep{MCCRACKEN12,MILVANG13} is able to search for potential LAE candidates at $z\sim8.8$ on 0.9 deg$^{2}$ on sky down to $1.5\times10^{-17}$erg~s$^{-1}$~cm$^{-2}$. Assuming this as limiting Ly$\alpha$ line flux and combined with our model from the LBG UV LF (optimistically assuming $X_{Ly\alpha}(z=8.8)=1$) it is unlikely that this survey will find LAEs at this redshift (expected counts are $0.6\pm0.3$). 
Likewise, with the same assumptions, Euclid \citep{LAUREIJS11} is not expected to find LAEs at $z>8$ with its spectroscopic configuration (1.1$\mu m$ - $2\mu m$, $3\times10^{-16}$erg/s/cm$^{2}$ on 20,000deg$^{2}$). On its proposed deep area (40deg$^{2}$) a flux limit of at least $3\times10^{-17}$erg/s/cm$^{2}$ must be reached to find one LAE at $z>8$.
Other space-based spectroscopic surveys like WISPs \citep{ATEK10} or 3D-HST \citep{BRAMMER12} using the HST grism G141, current flux limits around $5\times10^{-17}$erg~s$^{-1}$~cm$^{-2}$, and area of $600-800$ arcmin$^{2}$ need to be substantially (roughly 5 times) deeper to find LAEs at $z\sim8.8$.
Very deep small area blind imaging surveys with instruments on 8-10m telescopes such as HAWK-I ($7.5\arcmin \times 7.5\arcmin$) or MOSFIRE ($6.1\arcmin \times 6.1\arcmin$) must reaching flux limits of $5\times10^{-18}$erg~s$^{-1}$~cm$^{-2}$ in NB118 to pick up one LAE at $z\sim8.8$ on a total of $\sim10$ pointings.

\section{Conclusions}
We have presented follow-up observations on two bright LAE candidates at $z\sim7.7$ using MOSFIRE. We rule out any line emission at a level of several $\sigma$ for both objects. The limits inferred from these non-detections suggest a strong evolution of the LAE LF between $6 < z < 8$, consistent with what is seen in LBG samples.
We create an empirical model using the observed LBG UV continuum LFs and Ly$\alpha$ rest-frame equivalent width distributions to understand the interplay between LAE and UV continuum selected galaxies.
We find that our model and the observed LAE LF follow each other, but note a secondary effect which is due to a change in the EW$_{0}$ distribution of the galaxies as a function of redshift.
From this differential evolution and assuming two different models on Ly$\alpha$ absorption, we find consistent lower limits on the neutral hydrogen fraction at $z\sim7.7$ of 50-70\%.
Furthermore, we find a strong evolution in the Ly$\alpha$ optical depth at $z>6$ which can be characterized by $(1+z)^{2.2\pm0.5}$. All in all, our results are indicative of a continuation of strong evolution in the IGM beyond $z=7$.



\acknowledgments

We would like to acknowledge the support of the Keck Observatory staff who made these observations possible as well as B. Trakhtenbrot and W. Hartley for valuable discussions. We also thank Nick Konidaris for providing and supporting the MOSFIRE reduction pipeline and Gwen Rudie for providing the MOSFIRE exposure time calculator. AF acknowledges support from the Swiss National Science Foundation. AF also thanks Caltech for hospitality while this article was worked on.
The authors wish to recognize and acknowledge the very significant cultural role and reverence that the summit of Mauna Kea has always had within the indigenous Hawaiian community. We are most fortunate to have the opportunity to conduct observations from this mountain.



{\it Facilities:} \facility{Keck:I (MOSFIRE)}

\clearpage

\newpage

\newpage



\clearpage

\begin{deluxetable}{ccccccc}
\tabletypesize{\scriptsize}
\tablecaption{Large LAE surveys at $3 < z < 5$\label{tab:LF1}}
\tablewidth{0pt}
\tablehead{
\colhead{Redshift} & \colhead{phot. candidates} & \colhead{spec. confirmed / observed} & \colhead{spectr. fraction\tablenotemark{a}} & \colhead{Limits [AB]} & \colhead{Area\tablenotemark{b}} & \colhead{Ref}
}
\startdata
3.1 & 356 & 41/- & 12\% & $25.3$ (NB503) & $5\times0.2$&\citet{OUCHI08}\\
                
3.1 & 160 & - & 0\% & $25.4$ (NB5000) & $1\times0.28$ & \citet{GRONWALL07} \\
                
3.7 & 101 & 26/- & 26\% & $24.7$ (NB570) & $5\times0.2$&\citet{OUCHI08}\\
                
4.9 & 87 & - & 0\% & $26.0$ (NB711) & $1\times0.17$&\citet{OUCHI03}\\
\enddata
\tablenotetext{a}{Fraction of spectroscopically confirmed galaxies used in the analysis.}
\tablenotetext{b}{Given in deg$^{2}$.}
\end{deluxetable}

\clearpage
\begin{landscape}

\begin{deluxetable}{cccccccc}
\tabletypesize{\scriptsize}
\tablecaption{LAE surveys at $z\sim 5.7$, 6.6, 7.7, and 8.8\label{tab:LF2}}
\tablewidth{0pt}
\tablehead{
\colhead{Redshift} & \colhead{phot. cand.} & \colhead{spec. conf. / observed} & \colhead{spectr. fraction\tablenotemark{a}} & \colhead{Limits [AB]} & \colhead{Type of limit and aperture} & \colhead{Area\tablenotemark{b}} & \colhead{Ref}
}
\startdata
5.7 & 89 & 46/66 + 8\tablenotemark{c} & 55\% & $26.0$ (NB816) & 5$\sigma$, 2$\arcsec$ aperture & $1\times0.25$&\citet{KASHIKAWA11}\\

		5.7 & $\sim140$ & 87/140\tablenotemark{d} & 100\% & $25.3$ (NB816) & 5$\sigma$, 3$\arcsec$ aperture  & $7\times0.2$&\citet{HU10}\\

		5.7 & 401 & 17/29 & 4\% & $26.0$ (NB816) & 5$\sigma$, 2$\arcsec$ aperture & $5\times0.2$&\citet{OUCHI08}\\

		5.7 & 119 & - & 0\% & $25.1$ (NB816) &  5$\sigma$, 2$\arcsec$ aperture & $1\times1.95$ & \citet{MURAYAMA07}\\

		5.7 & 89 & 28/39 + 6/24 & 36\% & $26.6$ (NB816) &  3$\sigma$, 2$\arcsec$ aperture & $1\times0.2$&\citet{SHIMASAKU06}\\
		
		5.7 & 56$^{e}$ & 30/35 & 55\% & - &  & $\sim0.76$&\citet{MALHOTRA04}\\

\tableline

		6.6 & 207 (+58)\tablenotemark{e} & 16/24 (+ 16.22 + 1)\tablenotemark{e} & 13\% & $26.2$ (NB921) &  3$\sigma$, 2$\arcsec$ aperture & $5\times0.2$&\citet{OUCHI10}\\
                
                6.6 & 58 & 42/52 + 3\tablenotemark{f} & 74\% & $26.0$ (NB921) &  5$\sigma$, 2$\arcsec$ aperture & $1\times0.25$&\citet{KASHIKAWA11}\\
                
                6.6 & $\sim70$ & 30/70 & 100\% & $25.2$ (NB912) &  5$\sigma$, 3$\arcsec$ aperture & $7\times0.2$& \citet{HU10}\\
                
                6.6 & 61\tablenotemark{g} & 12/23 & 20\% & - &  & $\sim0.82$&\citet{MALHOTRA04}\\
                
\tableline        
                
		7.7 & 4 & 0/2 & 0\% & $22.4$ (UNB1056) & 50\% compl., auto aper. & $1\times0.2$& \citet{KRUG12}\\
                
		7.7 & 0 & - & - & $26.0$ (NB1060) & 5$\sigma$, $\sim1$\arcsec aperture & $3\times0.02$& \citet{CLEMENT12}\\
                
		7.7 & 7 & 0/5 & 0\% & $25.2$ (NB1060) & 4$\sigma$, 1.5\arcsec apert. (= 50\% compl.) & $1\times0.1$& \citet{HIBON10}\\
                 
		7.7 & 4 & - & 0\% & $22.5$ (UNB1063) & 50\% completenes & $1\times0.2$& \citet{TILVI10}\\                
                
\tableline
                
                8.8 & 13\tablenotemark{h} & 0/5\tablenotemark{h} & 0\% & $22.2$ (NB$_{J}$) & 5$\sigma$, 2$\arcsec$ aperture & $\sim 10$& \citet{MATTHEE14}\tablenotemark{i}\\  
                

\enddata
\tablenotetext{a}{Fraction of spectroscopically confirmed galaxies used in the analysis.}
\tablenotetext{b}{Given in deg$^{2}$.}
\tablenotetext{c}{20 in addition to \citet{SHIMASAKU06}.}
\tablenotetext{d}{Part of this sample is based on \citet{HU04}.}
\tablenotetext{e}{Based on \citet{KASHIKAWA06}.}
\tablenotetext{f}{28 in addition to \citet{KASHIKAWA06} and \citet{TANIGUCHI05}.}
\tablenotetext{g}{This sample is combined from different studies. Corrections for false detections are applied to the LFs. See \citet{MALHOTRA04} for more information.}
\tablenotetext{h}{Including 2 with J and K detections.}
\tablenotetext{i}{See also \citet{SOBRAL09,SOBRAL13}.}
\end{deluxetable}

\clearpage
\end{landscape}



\end{document}